\documentclass[
    ,final            
  ]
  {aipproc}

\layoutstyle{6x9}

\begin{document}

\title{A New Approach for Single Transverse-Spin Asymmetries from Twist-3 Soft-Gluon Mechanism}


\author{Yuji Koike}{
  address={Department of Physics, Niigata University,
Ikarashi, Niigata 950-2181, Japan}
}


\author{Kazuhiro Tanaka}{
  address={Department of Physics, 
Juntendo University, Inba-gun, Chiba 270-1695, Japan}
}

\begin{abstract}
A dominant QCD mechanism for the single transverse-spin asymmetry 
in hard processes 
is induced by the twist-3 quark-gluon correlations inside nucleon,
combined with the soft-gluonic poles to produce the interfering phase
for the associated partonic hard scattering. 
It is shown that
the corresponding
interfering amplitude
can be calculated entirely in terms of
the partonic Born cross section
which participates in the twist-2 cross section formula for the spin-averaged 
process.
\end{abstract}

\maketitle



We investigate the single-transverse spin asymmetry (SSA),
which is observed as the effect proportional 
to $\vec{S}_\perp \cdot(\vec{p} \times \vec{q})$
in the cross section for the scattering of transversely polarized proton with momentum $p$
and spin $S_\perp$,
off unpolarized particle with momentum $p'$, producing a particle with momentum $q$ 
which is observed in the final state.
Famous examples~\cite{ogawa} are $p^\uparrow p\to\pi X$, where the large asymmetry $A_N \sim 0.3$
was observed in the forward direction, 
and semi-inclusive deep 
inelastic scattering (SIDIS), $ep^\uparrow\to e\pi X$.
The Drell-Yan (DY) process, $p^\uparrow p\to \ell^+ \ell^- X$,
and the direct $\gamma$ production, $p^\uparrow p\to \gamma X$,
at RHIC, J-PARC, etc. are also expected to play important roles
for the study of SSA because those processes are clean 
in that 
fragmentation functions do not participate.

The SSA requires,
(i) nonzero $q_\perp$ originating 
from transverse motion
of quark or gluon; 
(ii) proton helicity flip; 
and (iii) 
interfering phase for the cross section. 
For processes with $q_\perp\sim \Lambda_{\rm QCD}$, all (i)-(iii) are generated 
from nonperturbative QCD mechanism
associated with the transverse-momentum-dependent parton distributions 
such as Sivers function~\cite{JQVY06,JQVY06DIS}.
On the other hand, for large $q_\perp \gg \Lambda_{\rm QCD}$,
perturbative effects also play essential roles to produce the SSA, in the framework of the
collinear factorization with twist-3 distributions:
In the DY pair production with $q_\perp \gg \Lambda_{\rm QCD}$, for example,
the large $q_\perp$ is provided by hard interaction
as the recoil from a hard final-state parton, as illustrated in Fig.~1.
Proton helicity flip is provided by the participation of the coherent, nonperturbative gluon
from the transversely polarized proton, the lower blob in Fig.~1; 
the lower part of Fig.~1 represents
the twist-3 quark-gluon correlation functions such as $G_{F}(x_1, x_2)$ 
with $x_1$ ($x_2$)
the lightcone momentum 
fraction of the quark leaving from (entering into) the proton~\cite{EKT}.
Due to the coupling of this coherent gluon,
some parton propagators in the partonic subprocess can be on-shell,
and this produces the imaginary phase as the pole contribution using
$1/(k^2 + i\varepsilon) = P(1/k^2) - i\pi\delta(k^2)$.
Depending on the resulting values of the coherent-gluon's momentum $k_g$, 
these poles are called soft-gluon pole (SGP) for $k_g=0$, and soft-fermion pole (SFP)
and hard pole (HP) for $k_g \neq 0$.


\begin{figure}
\includegraphics[height=.15\textheight]{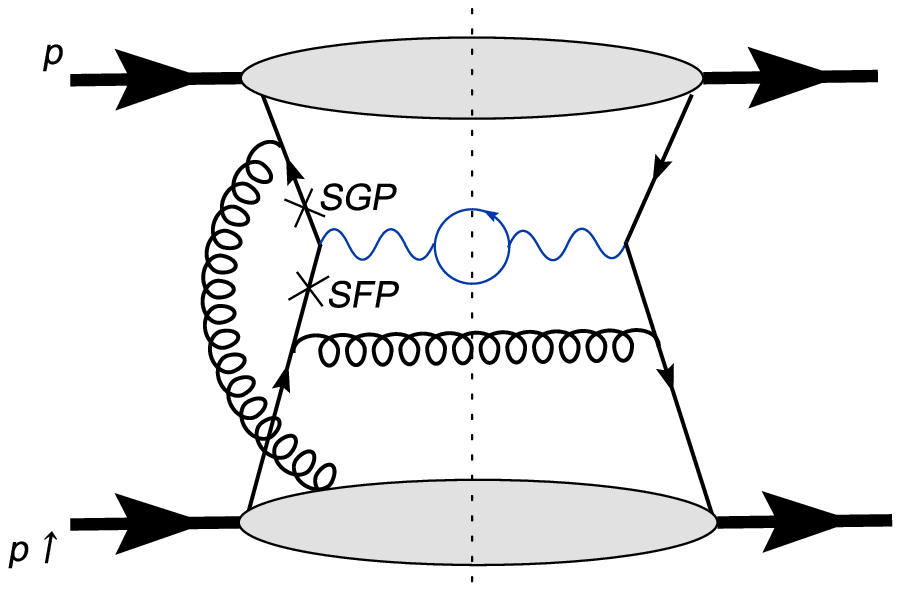}\hspace{1.4cm}
\includegraphics[height=.15\textheight]{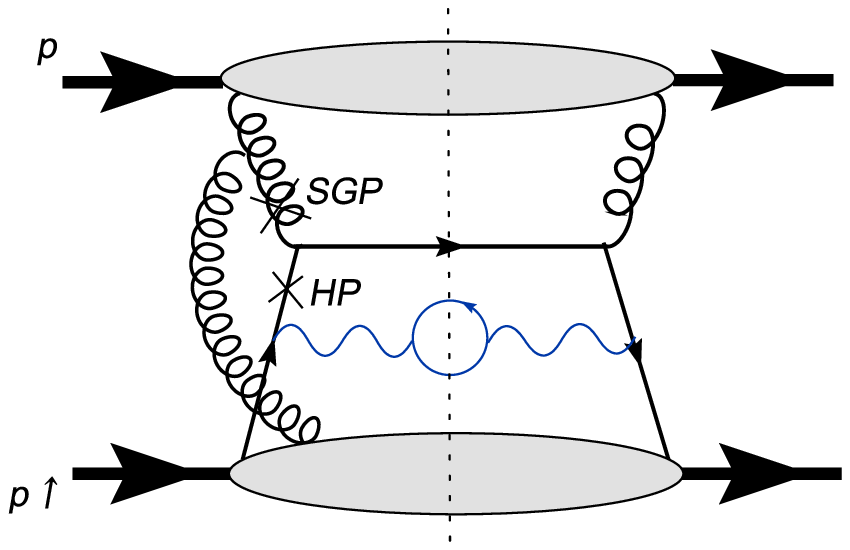}
\caption{Typical diagrams for the DY pair production with large $q_\perp \gg \Lambda_{\rm QCD}$
in the $q\bar{q}$-annihilation
and $qg$-scattering channels. The cross $\times$ denotes the parton propagator which gives the pole contribution.}
\end{figure}

Among these three-types of poles, the SGP plays somewhat distinct roles
compared with the other two, from theoretical as well as phenomenological view point
(see e.g. \cite{EKT,KT}). In particular, 
the SSA from the SGP contribution
is known to obey
remarkable simplicity: The partonic hard cross section 
associated with the ``derivative term'' ($\propto 
dG_F (x, x)/dx$) is identical,
up to certain kinematical and color factor, to that 
for 
the unpolarized 
twist-2 cross section 
as observed for direct $\gamma$ production~\cite{QS91}, DY process~\cite{JQVY06}, 
and SIDIS~\cite{EKT}. In this work we concentrate on the SGP contribution to SSA.

We have developed a systematic diagrammatic manipulation approach, 
and find that the SGP contributions from many Feynman diagrams 
are united into a derivative of the Born diagrams without coherent gluon line: 
The SSA for the DY process can be expressed as~\cite{KT} 
\begin{equation}
\frac{d\sigma^{\rm SGP}_{\mbox{\footnotesize tw-3}}}{[d\omega]}= 
\frac{\pi M_N}{2C_F}
\epsilon^{\sigma pnS_\perp} \sum_{j=\bar{q},g}B_j \int \frac{dx'}{x'}\int \frac{dx}{x} f_j(x') 
\frac{\partial H_{jq}(x', x)}{\partial (x' p^{\prime \sigma})} G_F^q(x,x),
\label{tw3formula}
\end{equation}
where $j= \bar{q}$ and $g$ represent  the ``$q\bar{q}$-annihilation'' 
and ``$qg$-scattering'' channels, respectively,
the sum over all quark and antiquark flavors, $q=u, \bar{u}, d, \bar{d}, \cdots$, is implicit for 
the index $q$, and 
$f_{j}(x')$ denotes the twist-2 parton distribution functions for the unpolarized proton.
$M_N$ is the nucleon mass representing nonperturbative scale
associated with the twist-3 correlation function $G_F^q(x,x)$ for the flavor $q$. 
$[d\omega]=dQ^2 dy d^2 q_\perp$ denotes the relevant differential elements 
with $Q^2 = q^2$ and $y$ the rapidity of the virtual photon.
The color factors are introduced as 
$B_q = 1/( 2N_c )$ and $-1/(2N_c )$ for quark and antiquark flavors, 
respectively,
$B_g = N_c/2$, and $C_F = (N_c^2-1)/(2N_c)$.
$H_{jq}(x', x)$ 
denote the partonic hard-scattering functions 
which participate in the unpolarized twist-2 cross section for 
DY process as
\begin{equation}
\frac{d\sigma^{\rm unpol}_{\mbox{\footnotesize tw-2}}}{[d\omega]}= 
\sum_{j=\bar{q},g}  \int \frac{dx'}{x'} \int \frac{dx}{x} 
f_j(x') H_{jq}(x', x) f_q(x).
\label{tw2formula}
\end{equation}
Namely, in order to obtain the explicit formula for the twist-3 SGP contributions to the SSA,
knowledge of the twist-2 unpolarized cross section is sufficient.

A proof of (\ref{tw3formula}) is given in \cite{KT}, and here we  
briefly sketch its main points: 
In the coupling of the coherent-gluon field $A^\mu$ to the partonic subprocess as in Fig.~1,
the scalar-polarized part, 
whose polarization vector is proportional to the gluon's momentum $k_g^\mu$, 
and the transversely-polarized part are relevant
up to the twist-3 accuracy in the Feynman gauge.
The coupling of the scalar-polarized part can be immediately manipulated using Ward identity, 
so that the gluon's interaction vertex, 
as well as the SGP from the parton propagator adjacent to it,
is disentangled.
On the other hand, the coupling of the transversely-polarized part to
the parton propagator can be decomposed into 
the eikonal vertex, combined with the eikonal propagator,
and the ``contact term'' containing the ``special propagator''~\cite{QS91}.
After this reduction,
the collinear expansion of the relevant diagrams 
in terms of the parton transverse momenta is quite straightforward,
and various contributions 
cancel with the corresponding contributions
from the ``mirror'' diagrams. 
The surviving twist-3 contributions are given
as the derivative of certain diagrammatic expression 
in terms of $k_{g\perp}$,
which can be compactly represented as (\ref{tw3formula})
in terms of the response of partonic Born subprocess
to the change of the transverse momentum carried by the parton inside unpolarized proton. 
Those manipulations can be performed for both $q\bar{q}$-annihilation and $qg$-scattering channels.
In particular, the three-gluon coupling relevant to the latter case can be treated,
without participation of the ghost-like gauge terms,
in exactly similar manner as the quark-gluon coupling for the former,
using the background field gauge~\cite{KT}.

When $Q^2 \rightarrow 0$, (\ref{tw3formula}) and (\ref{tw2formula})
reduce to the formulae for the direct $\gamma$ production. Also, ``crossing'' 
the initial unpolarized proton
into the final-state unpolarized hadron, and the virtual photon into the initial-state one
with $q^2 \rightarrow  -Q^2$, the DY process
can be formally transformed into the SIDIS. 
The proof discussed above is unaffected by 
the analytic continuation corresponding to this ``crossing transformation'', 
so that (\ref{tw3formula}), (\ref{tw2formula}) are also
applicable to the SIDIS with the appropriate substitutions.

Let us apply our ``master formula'' (\ref{tw3formula}) 
to obtain the explicit form for the SGP contribution to the SSA in the DY process.
The hard-scattering functions $H_{jq}(x', x)$
can be extracted
from the twist-2 factorization formula for the unpolarized DY process as
\begin{equation}
H_{jq} (x', x)= \frac{\alpha_{em}^2 \alpha_s e_q^2}{3\pi N_c s Q^2}
\widehat{\sigma}_{jq}(\hat{s}, \hat{t}, \hat{u}) 
\delta \left( \hat{s}+\hat{t}+\hat{u}-Q^2 \right),
\label{hdy}
\end{equation}
where $s=(p+p')^2$, and explicit form of $\widehat{\sigma}_{jq}(\hat{s}, \hat{t}, \hat{u})$ 
in terms of 
the partonic Mandelstam variables $\hat{s}, \hat{t}$ and $\hat{u}$
can be found in Eqs.~(12) and (14) in \cite{JQVY06}. The derivative in (\ref{tw3formula}) 
can be performed through that for the $\hat{u}$, and this may act on either 
$\widehat{\sigma}_{jq}$
or the delta function of (\ref{hdy}), where the latter case produces the derivative term with 
$dG_F (x, x)/dx$, as well as non-derivative term $\propto G_F (x, x)$,
by partial integration with respect to $x$. 
We get
\begin{eqnarray}
\lefteqn{
\frac{d\sigma^{\rm SGP, DY}_{\mbox{\footnotesize tw-3}}}{dQ^2 dy d^2 q_\perp}= 
\frac{\alpha_{em}^2 \alpha_s e_q^2}{3\pi N_c s Q^2}
\frac{\pi  M_N}{C_F} 
\epsilon^{pnS_\perp q_\perp}
\sum_{j=\bar{q}, g}B_{j}
\int \frac{dx'}{x'} \int \frac{dx}{x} \delta \left( \hat{s}+\hat{t}+\hat{u}-Q^2 \right)
f_j(x') 
}
\nonumber\\
&&\;\;\;\;\;\;\times
\left\{  \frac{\widehat{\sigma}_{jq}}{-\hat{u}}  x\frac{dG_F^q (x, x)}{dx}
+\left[\frac{\widehat{\sigma}_{jq}}{\hat{u}}  
-\frac{\partial \widehat{\sigma}_{jq}}{\partial \hat{u}}  
- \frac{\hat{s}}{\hat{u}}\frac{\partial  \widehat{\sigma}_{jq}}{\partial \hat{s}}  
-\frac{\hat{t}-Q^2}{\hat{u}}\frac{\partial  \widehat{\sigma}_{jq}}{\partial \hat{t}}  
\right] G_F^q (x, x)\right\}.
\label{dyf}
\end{eqnarray}
Substituting explicit form for $\widehat{\sigma}_{jq}(\hat{s}, \hat{t}, \hat{u})$,
this result completely coincides with that in \cite{JQVY06} obtained by 
direct evaluation of each Feynman diagram.
This result also explains why the partonic hard scattering functions associated 
with the derivative term
are directly proportional to those participating in the twist-2 unpolarized process,
$\widehat{\sigma}_{jq}$. Furthermore, our result reveals 
that the partonic hard-scattering functions associated 
with the non-derivative term are also
completely determined by $\widehat{\sigma}_{jq}$.
In the real-photon limit, $Q^2 \rightarrow 0$, (\ref{dyf}) gives 
the SSA for the direct $\gamma$ production (see \cite{JQVY06}).

We now use our master formula (\ref{tw3formula}) for the SIDIS, $ep^\uparrow\to e\pi X$.
We perform the substitutions corresponding to 
the crossing transformation discussed above:
$p' \rightarrow - P_{h}$, $x' \rightarrow 1/z$, $f_{\bar{q}}(x')\rightarrow D_q (z)$, 
$f_g(x')\rightarrow D_g (z)$, 
and $q^\mu \rightarrow -q^{\mu}$,
where $D_j(z)$ denote 
the twist-2 parton fragmentation functions for the final-state pion with momentum $P_h$,
and the new $q^{\mu}$ gives $Q^2 =- q^2$.
From (\ref{tw3formula}) and (\ref{tw2formula}), 
we get ($C_{q} \equiv B_{\bar{q}}, C_g \equiv B_g$)~\cite{KT}
\begin{equation}
\frac{d\sigma^{\rm SGP, SIDIS}_{\mbox{\footnotesize tw-3}}}{[d\omega]}= 
\frac{\pi M_N}{C_F z_f^2}
\epsilon^{pn S_\perp P_{h \perp}} \left. \frac{\partial}{\partial q_T^2}
\frac{d\sigma^{\rm unpol, SIDIS}_{\mbox{\footnotesize tw-2}}}{[d\omega]}
\right|_{f_q(x)\rightarrow G_F^q(x,x),\ D_j(z) \rightarrow C_j zD_j (z)},
\label{sidis}
\end{equation}
in a frame where the 3-momenta $\vec{q}$ and $\vec{p}$ of the
virtual photon and the transversely polarized nucleon are collinear along the $z$ axis.
$[d\omega]= dx_{bj}dQ^2 dz_f dq_T^2 d\phi$,
where, as usual, $x_{bj}={Q^2/ (2p\cdot q)}$, $z_f={p\cdot P_h / p\cdot q }$,
$q_T = P_{h\perp}/z_f$,
and $\phi$ is the azimuthal angle between the lepton and hadron planes. 
The twist-2 unpolarized cross section in the RHS of (\ref{sidis}) is known to have
several terms corresponding to different dependence on $\phi$,
which are proportional to $1, \cos \phi$, and $\cos 2\phi$, respectively (see \cite{ekt06}).
Performing the derivative in (\ref{sidis}) explicitly, the result
obeys exactly the same pattern as (\ref{dyf}) for both derivative and non-derivative terms.
Also, the result completely coincides, for all azimuthal dependence including those beyond the Sivers effect,
with the one that has been obtained recently in \cite{ekt06}
by direct evaluation of each Feynman diagram.

To summarize, we have discussed the twist-3 mechanism for the SSA arising from the SGP. 
We have developed a new approach that allows
systematic reduction of the coupling of the soft coherent-gluon and the associated 
pole contribution,
using Ward identities and decomposition identities for the interacting parton propagator, 
and derived 
the master formula which gives the twist-3 SGP contributions to the SSA
entirely in terms of the knowledge of the twist-2 factorization formula 
for the unpolarized cross
section. Our master formula is applicable to a range of processes, 
DY process, direct $\gamma$ production, SIDIS, and hopefully other processes.




\begin{theacknowledgments}
K.T. is 
supported by the Grant-in-Aid for Scientific Research No. C-16540266. 
\end{theacknowledgments}

\end{document}